# Order quantification of hexagonal periodic arrays fabricated by in situ solvent-assisted nanoimprint lithography of block copolymers

Claudia Simão,[a*] Worawut Khunsin,[a] Nikolaos Kehagias,[a] Mathieu Salaun,[b] Marc Zelsmann,[b] Michael A. Morris,[c,d] and Clivia M. Sotomayor Torres[a,e]

[a] Catalan Institute of Nanotechnology, Campus de la UAB, Barcelona 08193, Spain; [b]Laboratoire des Technologies de la Microélectronique, CNRS/UJF-Grenoble1/CEA LTM, 17 rue des Martyrs, 38000 Grenoble, France; [c] School of Chemistry and the Tyndall National Institute, UCC, Cork Ireland; [d] Centre for Research on Adaptive Nanostructures and Nanodevices, Trinity College Dublin, Ireland; [e] Catalan Institute of Research and Advanced Studies (ICREA), Barcelona 08010, Spain;

Directed self-assembly of block copolymer *polystyrene-b-polyethylene oxide* (PS-b-PEO) thin film was achieved by one-pot methodology of solvent vapor assisted nanoimprint lithography (SAIL). Simultaneous solvent-anneal and imprinting a PS-b-PEO thin film on silicon without surface pre-treatments yielded a 250 nm line grating decorated with 20 nm diameter nanodots array over a large surface area of up to 4" wafer scale. Grazing-incidence small-angle X-ray scattering (GISAXS) diffraction pattern showed the fidelity of the NIL stamp pattern replication and confirmed the periodicity of the BCP of 40 nm. The order of the hexagonally arranged nanodot lattice was quantified by SEM image analysis using the opposite partner method and compared to conventionally solvent-annealed block copolymer films. The imprint-based SAIL methodology thus demonstrated an improvement in ordering of the nanodot lattice of up to 50% and allows significant time and cost reduction in the processing of these structures.

Block copolymers (BCPs) are versatile materials and the most attractive bottom-up alternative to date for the fabrication of well-defined complex periodic structures with length scales of 3 to 100 nm.[33] Added by the ability to tailor their chemical structures for desired functionalities,[32] Self-assembled BCPs have been explored in a wide range of technological applications such as storage devices,[34, 35] solar cells,[6, 40] nano-electronics,[2, 47] low-$k$ dielectrics,[14, 27, 44] biochemical applications[41, 46, 51] and alternative lithography strategies to reduce the fabrication costs for patterning of sub-100 nm critical dimension highly demanded by the semiconductor industry.[7, 19, 20, 31, 43] Beside the demand on cost reduction, for an uptake by the industry, the fabrication technology must offer very precise control of the nanostructure's dimensions, regularity, registry, defect density and long-range order has to be achieved, a challenge which still remains for successful integration of block copolymer lithography at industrial scale.[4, 24, 25, 37, 45]

Nanoimprint lithography (NIL) is a low-cost, large-scale a soft lithography technique to fabricate sub-micrometer sized features in both two- and three- dimensions. .[13, 50] The top-down NIL technique has recently been explored as a tool to *direct* the self-assembly (DSA) of BCPs to achieve pattern density multiplication..[36, 38, 42] Furthermore, NIL provides an additional advantage by promoting long-range order if the stamp features are commensurate with the BCP periodicity ($L_0$),[21, 39] which is in turn directly related to the chains length of the block, prompting a physical confinement to the polymer chains by graphoepitaxy as a responsible mechanism.

According to Bates theory, the microphase segregation of block A and B of a BCP is thermodynamically driven by a high value of the Flory-Huggins parameter ($\chi$) that varies inversely with the temperature, which implies that low temperature formation processes afford advantage[3, 9] We note that an additional requirement to maintain an acceptable value of $\chi$ is to use BCPs with blocks of high chemical contrast.

$$\chi_{AB} = \frac{V_m (\delta_A - \delta_B)^2}{RT} \quad \text{(Eq. 1)}$$

However, conventional thermal NILs typically use high temperatures, above the individual BCP blocks glass transition temperatures ($T_g$), thus constraining the microphase separation of the BCP. In an effort to overcome this drawback, solvent vapor-assisted nanoimprint lithography (SAIL) is a promising means to achieve high order microphase segregation of BCPs by combining the bottom-up low-temperature approach of BCP DSA and micropatterning promotion of long-range order of NIL technique.[48]

In the present work, we propose the SAIL technique as a tool to direct the self-assembly of high $\chi$ BCPs to obtain low defect density microphase segregation (Figure 1a) A setup able to imprint up to 4" wafer size has been made for the purpose (Figure 1b). The BCP system used was a cylinder-forming *poly(styrene-b-ethylene oxide)* (PS-b-PEO) (Figure 1c), possessing a Flory-Huggins parameter $\chi_{PS-PEO}$ of 29.8/T-0.0029 near room temperature.[12], [30] To solvent anneal the PS-PEO film, a mixture of toluene and water was used to obtain cylinders aligned perpendicular to the substrate plane *without* any surface pre-treatment.[29, 30] Silicon stamps with 250 nm line patterns, 500 nm pitch and 50 nm height, were fabricated using conventional lithographic techniques.[38] The size of the trenches ($L_s$) is chosen to commensurate with the BCP periodicity,[28] where *$L_s \approx 6xL_0$*. PDMS replicas of the silicon stamps are used as the imprint mask inthe SAIL technique. A 1% (w/w) solution of PS-b-PEO in toluene was spin coated on a silicon substrate to obtain a thin film 34±2 nm thick as measured by ellipsometry. The BCP coated 4" wafer was subsequently placed inside the SAIL chamber and solvent vapors of toluene and water (1:1) were flown through at 2 ml/min for 10 minutes, after which the solvent vapor flow was stopped and the stamp was brought into contact with the sample. The pressure was maintained for 40 minutes during the imprint process and then a nitrogen flow was allowed through the chamber to demould the stamp.

The resulting sample exhibited a visual light diffraction, indicating a good pattern replication (see Supporting Information), which is confirmed by field emission scanning electron microscopy (FE-SEM) inspection as shown in Figure 2a. For comparison, a reference sample was made by conventional solvent annealing followed by an exposure to a saturated atmosphere of toluene and water for three hours at room temperature.

FE-SEM top-view images of the SAIL PS-PEO samples give clear evidence of the good replication of the PDMS NIL stamp in the BCP film over an area as large as several square millimeters (Figure 2.a). At higher magnifications, 40 nm periodically spaced dots of 20 nm diameter are observed in the residual layer within the trenches (Figure 2.b). These nanodot arrays provide evidence of the microphase segregation of PS-b-PEO, where the dots are micelles of PEO block in the PS matrix as schematically shown in Figure 2e. A cross-sectional TEM image (Figure 2C) suggests that the PEO cylinders in the mesas are oriented parallel to the substrate with the mesa thickness of 30 nm, which is in a good agreement with the starting film thickness. The residual layer is estimated to be about 10 nm thick. Time-of-flight secondary ion mass spectrometry (ToF-SIMS) analysis verifies the presence of the organic block copolymer film[5] and elemental depth profile analysis confirmed the average film thickness (See Supporting Information). We note that the rounded shape of the gratings mesas is related to the under-filling effect of the NIL stamp grooves (50 nm deep), and the smooth flat surface in the mesas is due to the presence of a PS block wetting layer.[38]

The homogeneity of the nanodot pattern throughout the imprinted area is remarkable. Indeed, a clear improvement in the order of the pattern compared to the same BCP system obtained by graphoepitaxy on nanopatterned substrates can readily be observed.[28] The SAIL technique required minutes rather than hours to achieve the BCP microphase segregation resulting in about a factor of three in time efficiency. By basing our approach on a one-step process without any surface pre-treatment, our work offer a distinct contribution to published works and recent efforts to obtain long-range order in BCP systems.

According to the Fredrickson and Bates work,[11] the diffusion parallel ($D_{par}$) and perpendicular ($D_{perp}$) to planes of the polymer chains are sensitive to the mechanism assumed for the Brownian motion of the chains. In particular, in non-tangled or pre-annealed polymers, as is the case of the solvent-swollen polymer, this means that the diffusion anisotropy defined by the ratio $D_{par}/D_{perp}$ will increase exponentially as the aspect ratio of the core block increases. Hamersky *et al.* have reported this ratio values as large as 40 in *poly(styrene-b-isoprene)* lamellae[16] and up to 80 in *poly-(ethyleneoxide-b-ethylethylene)* cylinders.[15] In the present work, the chosen solvent mixture results in PEO cylinders perpendicularly oriented with respect to the substrate.. The directional force applied by the NIL stamp creates a motion parallel to the polymer chains, as depicted by the blue arrows in Figure 1d. This motion is favorable to the diffusion of polymer chains in the residual layer and as such leads to the formation of an homogeneous nanodot array in the trenches throughout the imprinted area. In contrast, , the directional motion oriented perpendicular to the polymer chains occurs in the imprinted mesas and is illustrated with a purple arrow in Figure 1d. This motion hinders the polymer chain diffusion in the mesas and results in an incomplete alignment of the cylinders.

Grazing-incidence small-angle X-ray scattering (GISAXS)[8, 17, 18] has been one of the main techniques to characterize structural features of nanostructured polymer surfaces and thin films, capable of providing information on the nanometer scale along both lateral and vertical dimension over macroscopic regions. Utilizing an area detector, it is possible to extract information such as film thickness, particle geometry, and features of nanostructured surfaces. In an attempt to provide further evidence of structural features obtained above via SEM imaging, we performed GISAXS experiments using the Diamond synchrotron light source. The experimental geometry is shown in Figure 3a, where we define the scattering vector, $q = (q_x, q_y, q_z)$ with $q_x$ aligned along the groove direction, $q_y$ orthogonal to $q_x$ in the film plane, and $q_z$ pointing along the surface normal. The sample-to-detector distance was fixed at 3 m and the x-ray source is operated at 8 keV and has a wavelength of $\lambda = 1.55$ Å. The incident angle was set to 0.20 degrees to achieve total external reflection off the silicon substrate, but allows full penetration of the BCP film.

Figure 3b shows the measured GISAXS pattern (same as that shown in Figure 3a) along with two profile cuts: one along $q_z$ direction and the other along $q_y$ direction. As a scattering technique, GISAXS transforms structural parameters in real space into the reciprocal space. For a line grating with the groove direction aligned along the incident beam path (i.e. Figure 3a), GISAXS pattern is made up of vertical rods, i.e. Bragg rods, as a result of intersection between the Ewald sphere and the reciprocal space representation of the grating lattice.[17, 18] These Bragg rods, which are equally spaced along the in-plane scattering vector $q_y$, is directly related to the lattice spacing in real-space through $d = 2\pi/\Delta q_y$, where $d$ and $\Delta q_y$ are the period between any two real-space lattice points and reciprocal-space Bragg rods, respectively. As observed in Figure 3b, our grating structure results in GISAXS pattern composed principally of vertical rods, which are equally spaced at an interval of $\Delta q_y = 0.012$ nm$^{-1}$. The corresponding real-space period is thus, $d = 2\pi/\Delta q_y = 515$ nm, which is in good agreement with the stamp pitch and the SEM observation mentioned above. It is worth noting that, the reciprocal and the corresponding real-space representations are essentially a Fourier pair of one another. Indeed, Wernecker and coworkers[49] have recently proposed a complementary structural analysis method based on Power Spectral Density (PSD) analysis obtained via Fourier Transform (FT) of the experimental scattering data. The authors demonstrated that the method provides a convenient way to extract parameters such as grating period, groove width and line height. Applying discrete FT to a profile cut at $q_z = 0.409$ nm$^{-1}$ and plot the resulting PSD profile versus the real-space correlation length, i.e., grating period $\Lambda$, we have the curve shown in Figure 3c. The two principal peaks in the curve corresponds to the real-space periods of 214 nm and 470 nm, which can be conveniently assigned to the grating line width and period, respectively.

A similar procedure is applied to the out-of-plane scattering profile, i.e., along $q_z$ direction (blue curve) and provides structural information in the vertical direction.[49] Figure 3d shows the resulting PSD profile consisting of a main peak at 26.18 nm, which is in reasonably close agreement with the thickness of the mesas measured by TEM.

The diffuse Debye-Scherrer ring in the GISAXS pattern (Figure 3a) is due to the diffraction from the same family of lattice planes (i.e. same lattice spacing) of the hexagonal array of horizontally aligned cylinders, but with different orientations along the cylinder axis.[26] Correspondingly, we attributed the observed ring to the disordered lattice in the mesas. The in-plane component, $q_y$, of the ring ($q_y$ = 0.15 nm$^{-1}$) corresponds to a real-space period of 41.8 nm of the hexagonal lattice of horizontally aligned cylinders in the mesas, which is in good agreement with the values obtained above from the cross-sectional TEM image inspection. (see Figure 2c) We note, however, that the diffraction pattern pertaining to the hexagonal array of vertical pillars in the residue layer is not observed in our experiment. In order to analyze the GISAXS pattern further, we simulated the GISAXS pattern using the open source software package *Scatter*.[10] The simulated structure is composed of a hexagonal array of vertical pillars having the same structural parameter as determined experimentally from SEM images (Figure 2): height = 10 nm, diameter = 20 nm, lattice spacing = 40 nm. The simulated pattern is shown in Figure 3e and 3f. GISAXS intensity is proportional to the product of the structure factor S(q), which contains information on the lattice arrangement, orientation, dimension and symmetry, and the form factor P(q), which contains information on the scattering object such as size, shape and orientation in the lattice. Mapping information in real-space into the reciprocal space,

GISAXS experiment essentially transforms a convolution in real space to a multiplication in the reciprocal space. For a two-dimensional hexagonal lattice, such a transformation results in the same lattice configuration, but rotated by 30 degrees with respect to the original lattice. In other words, one can analyze the structure factor S(q) (i.e. interference function) based on pair correlation function analysis of the real-space lattice. The vertical black lines shown in Figure 3f superimposed onto the zoom-in of the calculated GISAXS pattern are the result from the analysis of pair correlation function. The peak positions are in excellent agreement with the simulated GISAXS pattern. We note that the experimental GISAXS pattern shown in Figure 3a, b and the calculated GISAXS pattern of a hexagonal array of vertical pillar mimicking the patterns observed in the residue layer do not share common features. As such, information regarding the pattern in the residue layer cannot be extracted based on the current GISAXS experiment, due to the following difficulties: first, GISAXS signal coming from the residual layer is weak compared to that from the mesa, which is likely due to a factor of three times thinner film, thus less material in the residual layer.; and second, the hexagonal lattice in the mesas has the same real-space period as horizontally aligned cylinders in the mesas. This means that both patterns will produce the scattering peaks at the same $q_y = 0.17$ nm$^{-1}$. As such, the diffraction pattern resulting from the much weaker residual layer may be obscured by a much stronger signal from the mesas.

In an attempt to achieve quantitative information on the order of the hexagonal nanodots array we applied our recently developed image analysis technique,[23] which has been shown to be both robust and accurate compared to the commonly used Fourier Transform (FT) technique, and translational correlation function and Bond-orientation correlation function methods. In addition, our technique is well-suited for cross-sample comparison, particularly in the case of the structures investigated here where edge and binning effect will severely affect FT and correlation function analysis. In brief, our quantification technique utilizes the rotational symmetry of the PS-b-PEO hexagonal nanodots array and calculates the probability of finding an 'opposite partner', i.e. a centrally symmetric partner around a chosen central feature. Depending on the length scale of interest, given by the parameter *r* (nm), and the tolerance in the exact location of the feature center, given by the parameter $\varepsilon$ (nm) (see the inset to Figure 4a), the probability of finding an opposite partner and thus the quality of the structural order of the array is given by Equation 3

$$p(r) = \frac{\sum_{A \neq B, C} X_r(\vec{AB}) X_s(\vec{AB} + \vec{AC})}{\sum_{A \neq B} X_r(\vec{AB})} \quad \text{(Eq.3)}$$

where $\vec{AB}$ is the vector from the center of sphere A to the center of sphere B (Inset in Figure 4b) and the characteristic function $X_{r,\varepsilon}(\vec{R})$ is defined in Equation 4 as

$$X_{r,\varepsilon}(\vec{R}) = \begin{cases} 1 & if \ |\vec{R}| < r, \varepsilon \\ 0 & otherwise \end{cases} \quad \text{(Eq. 4)}$$

To obtain global order quantification, the analysis is applied to all the features in the SEM images to obtain the weighted average as described in Equation 5:

$$p(r) = \frac{\sum_A N_A(r) p_A(r)}{\sum_A N_A(r)} \quad \text{(Eq.5)}$$

where $p_A(r)$ represents the local regularity measured at a chosen feature *A* and $N_A(r)$ is the total number of features (excluding the central feature) within the length scale of interest *r*. We applied the technique to SEM images similar to that shown in Figure 2, the results of which are presented in Figure 4a), showing a comparison between arrays of PS-b-PEO microphase segregation organized in grooves by SAIL and those obtained by conventional solvent annealing on a flat substrate. Plotting the regularity measure *p(r)* as a function of tolerance parameter $\varepsilon$, it is evident that the arrays obtained by SAIL possess better order than those obtained on flat substrates. Specifically, given an uncertainty in the feature position of 5 nm, i.e., 0.25*d*, where *d* is the diameter of the cylinder, the SAIL technique yields a nanostructure array with 25% improvement in the ordering compared to those organized on a flat substrate by the conventional solvent annealing technique (*p(r)* of 0.5 vs 0.1). By allowing a larger tolerance parameter reaching 0.5d (i.e. 10 nm), larger order improvement of 50% is observed in the sample prepared by the SAIL technique.

On a final note, we would like to point out that in the case of self-assembly of block copolymer in confined environment, competition exists between bulk interaction, which regulates the formation of lamellae or cylinders with a natural period $L_0$, and surface interaction with the walls, which leads to preferential attraction of one block copolymer component to the interfaces. The outcome is the observation of perturbed lattice period *L'*, which can be smaller (compressive) or larger (tensile) than $L_0$. The oscillation between compressive and tensile strain becomes less pronounced for larger separation between the confining walls because the mismatch between the size of the trench and integer multiple of lattice period of the polymer chain is distributed among all the periods. It has been shown that tension is easier to accommodate than compression,[1, 22] which is also observed in our case. Specifically, we calculate normalized strain according to the formula: $\varepsilon = (Y-L_0)/L_0$, where Y is the average row-to-row spacing of block copolymer self-assembled inside the trench and $L_0$ is the natural period obtained from block copolymer self-assembled on flat substrate, i.e. without any confining walls.

Based on the obtained SEM images, we estimated the tensile strain observed in our experiment to be about 10%, which is in a very good agreement with previously reported values.[1, 22]

In summary, the SAIL technique developed here offers an exciting and straightforward approach for realizing large-area nanodots arrays in a one-step low-cost process. We demonstrate that the PS-PEO microphase segregation yields NIL density multiplication by direct imprinting of the BCP film with NIL-patterned mask.. The PEO cylinders in the residual layer are oriented homogenously perpendicular to the substrate in a dense hexagonal array throughout the imprinted area. In the mesas, the cylinders are mainly oriented parallel to the substrate with a certain degree of misalignment. Such alternating orientation can result in a formation of a complex patterned surface such as superlattices. The disorder of the features in the mesas is correlated with an anisotropic diffusion of the BCP chains upon an applied force and is currently being investigated as means to achieve control over the micro phase segregation effect leading, for example, to the realization of aligned nanowire arrays in the mesas. GISAXS measurements give the line grating width, height, and periodicity of BCP microphase segregation in the mesas in close agreement with the NIL stamp dimensions and the data obtained from SEM inspection, and provide a direct evidence of high quality stamp replication in the BCP film. In addition, we prove that the order of the nanodot array in the residual layer is improved by 50% if the self-assembly is directed by SAIL compared to conventional solvent annealing. Using the image analysis, it is possible to extract structural information and analysis of corrugated surfaces with alleviated dimensions, which is otherwise not easily obtained solely based on commonly used GISAXS experiments.

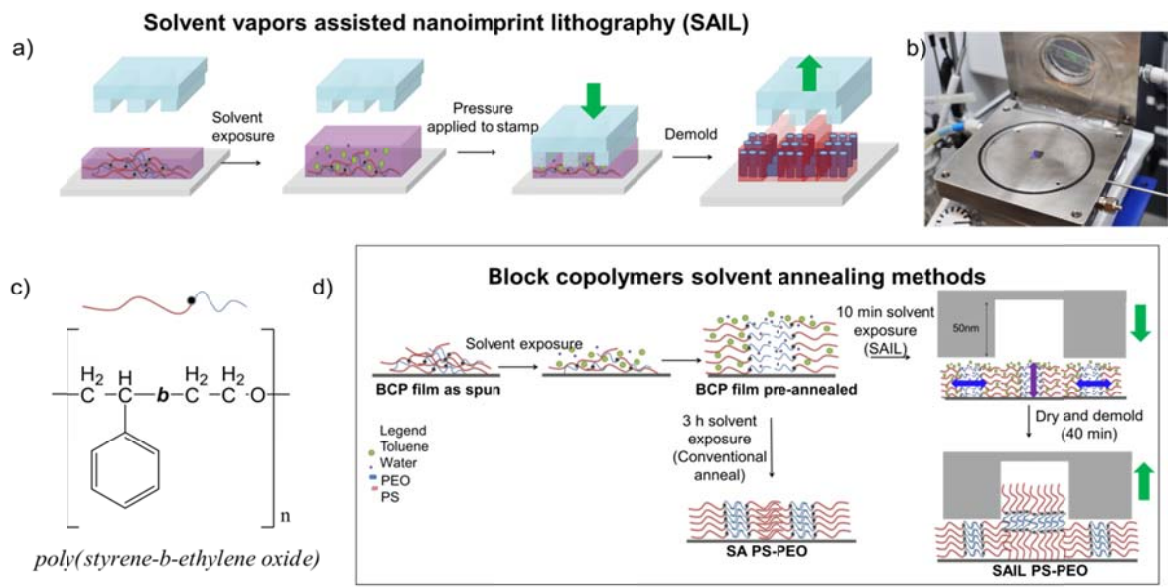

**Figure 1.** a) Process workflow of solvent-assisted nanoimprint lithography (SAIL). b) A photograph of the in-house made SAIL chamber. c) Structural formula of *poly(styrene-b-ethylene oxide)* block copolymer. d) Schematic representation of block copolymer film solvent annealing by conventional solvents exposure or solvent vapors assisted NIL. Green arrows indicate NIL stamp movement direction; blue and purple arrows indicate polymer chain movement direction during the imprint process.

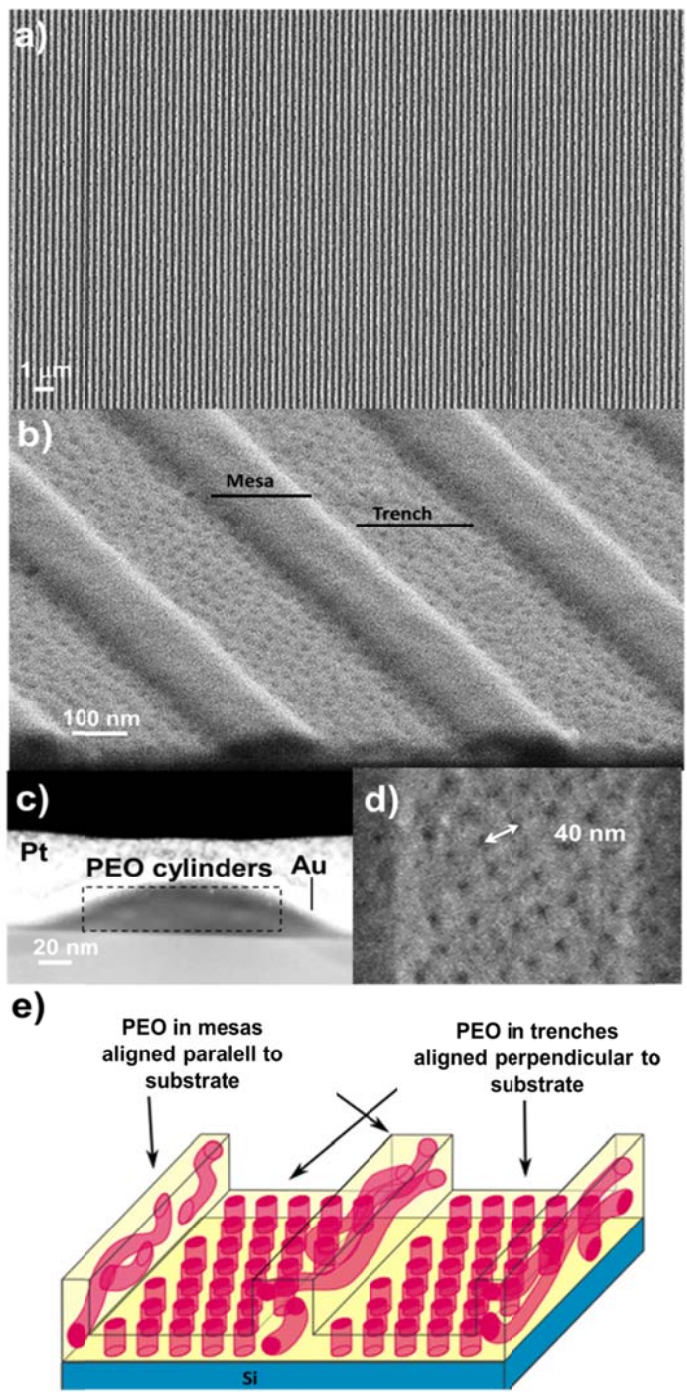

**Figure 2.** a) FE-SEM images of PS-b-PEO film imprinted in SAIL setup show homogeneous imprints of 250 nm lines over a large area. b) FE-SEM cross section image from the sample showing the nanodot array in the residual layer (10 nm thickness) and gratings height of approximately 30 nm. c) Dark-field TEM image of a FIB lamella of SAIL sample, showing PEO cylinder aligned horizontally parallel to the substrate. d) Close image of 20 nm dots with periodicity of 40 nm. e) Schematic representation of the PEO cylinders (in red) orientation in the PS matrix (in yellow).

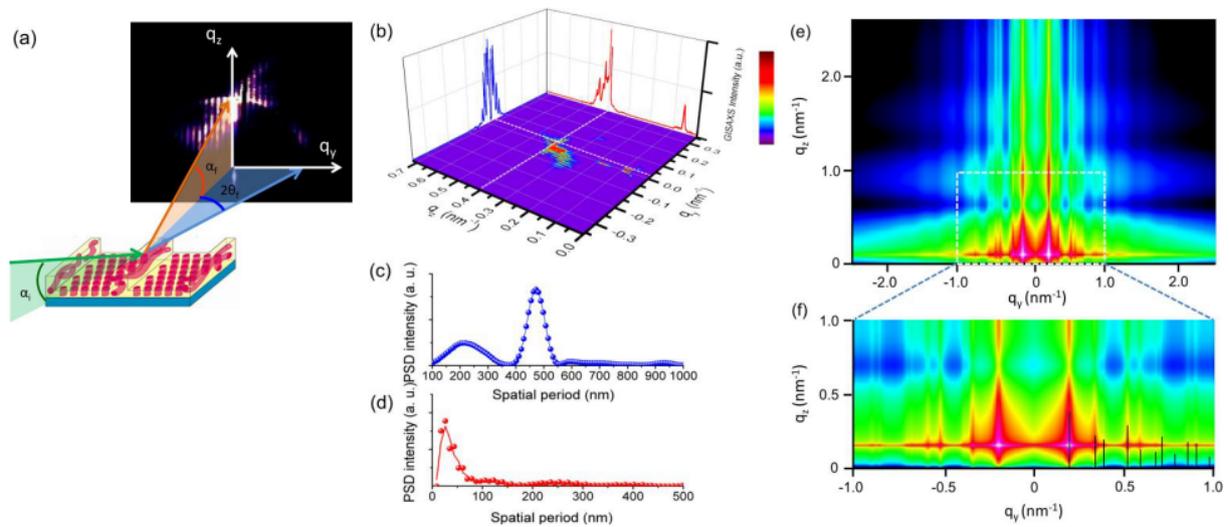

**Figure 3.** (a) GISAXS experimental geometry: Incident beam (green arrow), out-of-plane (orange) and in-plane (blue) scattering beams, where $α_i$, $α_f$, and $2θ_f$ denote the angle of incidence, the out-of-plane and in-plane exit angles, respectively. (b) A contour plot of the measured GISAXS intensity map as a function of $q_y$ (in-plane) and $q_z$ (out-of-plane) scattering vectors. Profile cuts are obtained at $q_z$ = 0.409 (blue curve) and $q_y$ = 0.032 (red curve) as indicated by the dashed white lines, respectively. (c) and (d) PSD of scattering profiles shown in (b) as a function of spatial period, i.e., characteristic scattering length. (e) Simulated GISAXS scattering pattern for an hexagonal array of vertically aligned cylinder obtained using Scatter software.[10] See

text for details. (f) a zoom-in of the panel (e). The vertical black lines are calculated pair correlation function of a two-dimensional hexagonal lattice.

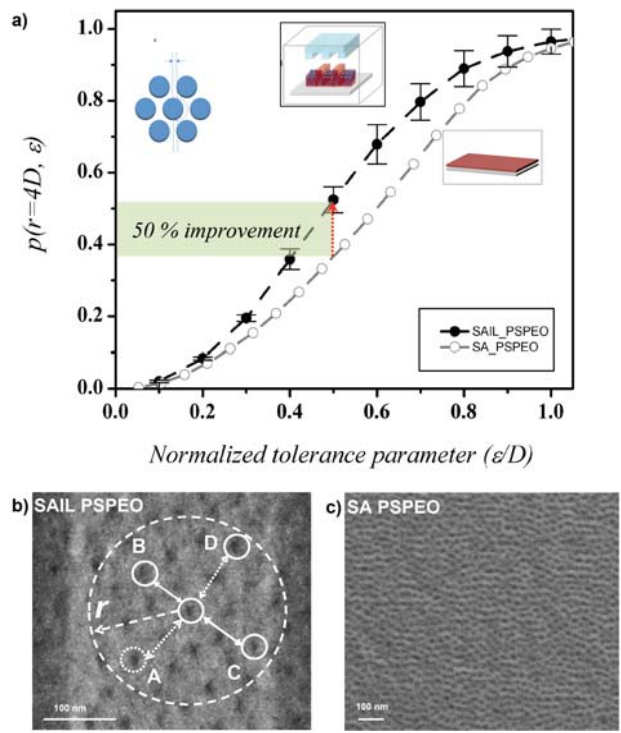

**Figure 4.** (a) Order quantification, p(r), of the PS-b-PEO films annealed by (b) SAIL or (c) conventional solvent annealing. Inset to (a): a schematic representation of a tolerance parameter $\varepsilon$ in hexagonal packing. Inset to (b): a schematic representation of order quantification algorithm.

AUTHOR INFORMATION

**Corresponding Author**

claudia.simao@icn.cat

**Funding Sources**

The research leading to these results received funding from the European Union FP7 under the project LAMAND (grant agreement n° 245565), NANOFUNCTION (grant agreement no. 257375, FP7-ICT-2009-5) and by the Spanish Ministry of Economics and Competitiveness under project TAPHOR contract no. MAT2012-31392 (Plan Nacional de I + D + I (2008-2011).

ACKNOWLEDGMENT

The contents of this work are the sole responsibility of the authors. The authors wish to thank Dr. P. Mokarian-Tabarian for BCP solvent annealing discussions, Dr. Ben O'Driscoll and Beamline I07 at Diamond synchrotron light source for GISAXS measurements.

ABBREVIATIONS

BCP, block copolymer; DSA directed self-assembly; NIL, nanoimprint lithography; PS-b-PEO, poly(styrene-block-ethylene oxide); PS, polystyrene; PEO, polyethylene oxide; SAIL, solvent vapors assisted nanoimprint lithography; SA, solvent annealing; GISAXS, grazing incident small angle X-ray scattering.

## Supporting Information

Figure S1. Picture of SAIL-imprinted PS-PEO thin film on a 4" wafer substrate, cut to fit the stamp. Some particles are seen trapped in the film.

Figure S2. Positive ion molecular depth profiles of PS-b-PEO (42k-11.5k) diblock copolymer film after annealing in SAIL process. Analysis ion beam = 25 keV $Bi^+$; sputter ion beam = 20 keV $C60^{++}$.

REFERENCES

[1] Asbahi M, Lim K T P, Wang F, Duan H, Thiyagarajah N, Ng V and Yang J K W 2012 Directed Self-Assembly of Densely Packed Gold Nanoparticles *Langmuir* **28** 16782-7
[2] Bai J, Zhong X, Jiang S, Huang Y and Duan X 2010 Graphene nanomesh *Nature Nanotechnology* **5** 190-4
[3] Bates F S 1991 *Science* **251** 898
[4] Borah D, Shaw M T, Rasappa S, Farrell R A, O'Mahony C, Faulkner C M, Bosea M, Gleeson P, Holmes J D and Morris M A 2011 Plasma etch technologies for the development of ultra-small feature size transistor devices *J. Phys. D: Appl. Phys.* **44**

[5] Borah D, Simao C D, Senthamaraikannan R, Rasappa S, Francone A, Lorret O, Salaun M, Kosmala B, Kehagias N, Zelsmann M, Sotomayor-Torres C M and Morris M A Soft-graphoepitaxy using nanoimprinted polyhedral oligomeric silsesquioxane substrates for the directed self-assembly of PS-b-PDMS *Eur. Polym. J.* **49** 3512–21

[6] Crossland E J W, Nedelcu M, Ducati C, Ludwigs S, Hillmyer M A, Steiner U and Snaith H J 2009 Block copolymer morphologies in dye-sensitized solar cells: Probing the photovoltaic structure-function relation *Nano Lett.* **9** 2813-9

[7] Cheng J Y, Ross C A, Chan V Z H, Thomas E L, Lammertink R G H and Vancso G J 2001 Formation of a cobalt magnetic dot array via block copolymer lithography *Adv. Mater.* **13** 1174-8

[8] Darhuber A A, Bauer G, Wang P D and Torres C M S 1998 Shear strains in dry etched GaAs/AlAs wires studied by high resolution xray reciprocal space mapping *Journal Applied Physics* **83** 126-31

[9] Farrell R A, Fitzgerald T G, Borah D, Holmes J D and Morris M A 2009 Chemical Interactions and their role in the microphase separation of block copolymer thin films. *International Journal of Molecular Science* **10** 3671-712

[10] Förster S, Apostol L and Bras W 2010) Scatter: software for the analysis of nano- and mesoscale small-angle scattering *J. Appl. Crystallogr.* **43** 639-46

[11] Fredrickson G H and Bates F S 1996 Dynamics of block copolymers: Theory and Experiment *Annu. Rev. Mater. Sci.* **26** 501-50

[12] Frielinghaus H, Hermsdorf N, Almdal K, Mortensen K, Messe L, Corvazier L, Fairclough J P A, Ryan A J, Olmsted P D and Hamley I W 2001 Micro- vs. macro-phase separation in binary blends of poly(styrene)-poly(isoprene) and poly(isoprene)-poly(ethylene oxide) diblock copolymers *Europhysics Letters* **53** 680-6

[13] Gates B D, Xu Q, Stewart M, Ryan D, Willson C G and Whitesides G M 2005 New approaches to nanofabrication: Molding, printing, and other techniques *Chem. Rev.* **105** 1171-96

[14] Ge J and Yin Y 2011 Responsive photonic crystals *Angew. Chem. Int. Ed.* **50** 1492-522

[15] Hamersky M W, Hillmyer M A, Tirrell M, Bates F S, Lodge T P and Meerwall E D v 1998 *Macromolecules* **31** 5363

[16] Hamersky M W, Tirrell M and Lodge T P 1998 Anisotropy of Diffusion in a Lamellar Styrene-Isoprene Block Copolymer *Langmuir* **14** 6974

[17] Hlaing H, Lu X, Hofmann T, Yager K G, Black C T and Ocko B M 2011 Nanoimprint-Induced Molecular Orientation in Semiconducting Polymer Nanostructures *ACS Nano* **5** 7532–8

[18] Hong S W, Voronov D L, Lee D H, Hexemer A, Padmore H A, Xu T and Russell T P 2012 Controlled Orientation of Block Copolymers on Defect-Free Faceted Surfaces *Adv. Mater.* **24** 4278–83

[19] Huo F, Zheng Z, Zheng G, Giam L R, Zhang H and Mirkin C A 2008 Polymer pen lithography *Science* **321** 1658-60

[20] Ito T and Okazaki S 2000 Pushing the limits of lithography *Nature* **406** 1027-31

[21] Jeong S J and Kim S O 2011 Ultralarge-area block copolymer lithography via soft graphoepitaxy *J. Mater. Chem.* **21** 5856-9

[22] Kellogg G J, Walton D G, Mayes A M, Lambooy P, Russell T P, Gallagher P D and Satija S K 1996 Observed Surface Energy Effects in Confined Diblock Copolymers *Phys. Rev. Lett.* **76** 2503

[23] Khunsin W, Amann A, Kocher-Oberlehner G, Romanov S G, Pullteap S, Seat H C, O'Reilly E P, Zentel R and Sotomayor Torres C M 2012 Noise-assisted crystallization of opal films *Adv. Funct. Mater.* **22** 1812-21
[24] Kikitsu A, Maeda T, Hieda H, Yamamoto R, Kihara N and Kamata Y 2013 5 Tdots/in2 bit patterned media fabricated by a directed self-assembly mask *IEEE Transactions on Magnetics* **49** 693-8
[25] Lan H and Ding Y 2012 Ordering, positioning and uniformity of quantum dot arrays *Nano Today* **7** 94-123
[26] Lee B, Park I, Yoon J, Park S, Kim J, Kim K-W, Chang T and Ree M 2005 In-situ Grazing Incidence Small Angle X-ray Scattering Studies on Nanopore Evolution in Low-k Organosilicate Dielectric Thin Films *Macromol.* **38** 4311-23
[27] Miyake G M, Piunova V A, Weitekamp R A and Grubbs R H 2012 Precisely tunable photonic crystals from rapidly self-assembling brush block copolymer blends *Angew. Chem. Int. Ed.* **51** 11246-8
[28] Mokarian-Tabari P, Collins T W, Holmes J D and Morris M A 2011 Cyclical "flipping" of morphology in block copolymer thin films *ACS Nano* **5** 4617-23
[29] Mokarian-Tabari P, Collins T W, Holmes J D and Morris M A 2012 Brushless and Controlled Microphase Separation of Lamellar Polystyrene-b-Polyethylene Oxide Thin Films for Block Copolymer Nanolithography *JOURNAL OF POLYMER SCIENCE: PART B: POLYMER PHYSICS* **50** 904-9
[30] Mokarian-Tabari P, Collins T W, Holmes J D and Morris M A 2012 Brushless and controlled microphase separation of lamellar polystyrene-b-polyethylene oxide thin films for block copolymer nanolithography *J. Polym. Sci., Part B: Polym. Phys.* **50** 904-9
[31] Moonen P F, Yakimets I and Huskens J 2012 Fabrication of transistors on flexible substrates: From mass-printing to high-resolution alternative lithography strategies *Adv. Mater.* **24** 5526-41
[32] Nie Z and Kumacheva E 2008 Patterning surfaces with functional polymers *Nat. Mater.* **7** 277-90
[33] Onses M S, Song C, Williamson L, Sutanto E, Ferreira P M, Alleyne A G, Nealey P F, Ahn H and Rogers J A 2013 Hierarchical patterns of three-dimensional block-copolymer films formed by electrohydrodynamic jet printing and self-assembly *Nat Nano* **8** 667-75
[34] Park M, Harrison C, Chaikin P M, Register R A and Adamson D H 1997 Block copolymer lithography: Periodic arrays of ~10 11 holes in 1 square centimeter *Science* **276** 1401-4
[35] Park S, Dong H L, Xu J, Kim B, Sung W H, Jeong U, Xu T and Russell T P 2009 Macroscopic 10-terabit-per-square-inch arrays from block copolymers with lateral order *Science* **323** 1030-3
[36] Park S M, Liang X, Harteneck B D, Pick T E, Hiroshiba N, Wu Y, Helms B A and Olynick D L 2011 Sub-10 nm nanofabrication via nanoimprint directed self-assembly of block copolymers *ACS Nano* **5** 8523-31
[37] Ramanathan M, Darling S B and Mancini D C 2011 Block copolymer lithography as a facile route for developing nanowire-like arrays *Advanced Science Letters* **4** 437-41
[38] Salaün M, Kehagias N, Salhi B, Baron T, Boussey J, Sotomayor Torres C M and Zelsmann M 2011 Direct top-down ordering of diblock copolymers through nanoimprint lithography *J. Vac. Sci. Technol. B* **29** 06F208

[39] Salaun M, Zelsmann M, Archambault S, Borah D, Kehagias N, Simao C, Lorret O, Shaw M, Sotomayor Torres C M and Morris M A 2013 Fabrication of highly ordered sub-20 nm silicon nanopillars by block copolymer lithography combined with resist design *Journal of Materials Chemistry C* **1** 3544–50

[40] Seo J H, Gutacker A, Sun Y, Wu H, Huang F, Cao Y, Scherf U, Heeger A J and Bazan G C 2011 Improved high-efficiency organic solar cells via incorporation of a conjugated polyelectrolyte interlayer *J. Am. Chem. Soc.* **133** 8416-9

[41] Shin D O, Jeong J R, Han T H, Koo C M, Park H J, Lim Y T and Kim S O 2010 A plasmonic biosensor array by block copolymer lithography *J. Mater. Chem.* **20** 7241-7

[42] Simão C, Francone A, Borah D, Lorret O, Salaun M, Kosmala B, Shaw M T, Dittert B, Kehagias N, Zelsmann M, Morris M A and Sotomayor-Torres C M 2012 Soft graphoepitaxy of hexagonal PS-b-PDMS on nanopatterned POSS surfaces fabricated by nanoimprint lithography *J. Photopolym. Sci. Technol.* **25** 239-44

[43] Somervell M, Gronheid R, Hooge J, Nafus K, Delgadillo P R, Thode C, Younkin T, Matsunaga K, Rathsack B, Scheer S and Nealey P 2012 Comparison of directed self-assembly integrations(vol 8325)

[44] Son J G, Hannon A F, Gotrik K W, Alexander-Katz A and Ross C A 2011 Hierarchical nanostructures by sequential self-assembly of styrene-dimethylsiloxane block copolymers of different periods *Adv. Mater.* **23** 634-9

[45] Thébault P, Niedermayer S, Landis S, Chaix N, Guenoun P, Daillant J, Man X, Andelman D and Orland H 2012 Tailoring nanostructures using copolymer nanoimprint lithography *Adv. Mater.* **24** 1952-5

[46] Uehara H, Kakiage M, Sekiya M, Sakuma D, Yamonobe T, Takano N, Barraud A, Meurville E and Ryser P 2009 Size-selective diffusion in nanoporous but flexible membranes for glucose sensors *ACS Nano* **3** 924-32

[47] Usta H, Risko C, Wang Z, Huang H, Deliomeroglu M K, Zhukhovitskiy A, Facchetti A and Marks T J 2009 Design, synthesis, and characterization of ladder-type molecules and polymers. air-stable, solution-processable n-channel and ambipolar semiconductors for thin-film transistors via experiment and theory *J. Am. Chem. Soc.* **131** 5586-608

[48] Voicu N, Ludwigs S, Crossland E J W, Andrew P and Steiner U 2007 Solvent-Vapor-Assisted Imprint Lithography *Adv. Mater.* **19** 757-61

[49] Wernecke J, Scholze F and Krumrey M 2012 Direct structural characterisation of line gratings with grazing incidence small-angle x-ray scattering *Rev. Sci. Instrum.* **83** 103906

[50] Zankovych S, Hoffmann T, Seekamp J, Bruch J U and Torres C M S 2001 *Nanotechnology* **12** 91

[51] Zhang Z B, Yuan S J, Zhu X L, Neoh K G and Kang E T 2010 Enzyme-mediated amperometric biosensors prepared via successive surface-initiated atom-transfer radical polymerization *Biosens. Bioelectron.* **25** 1102-8